\documentclass[aps, pra, twocolumn, floatfix, nobalancelastpage, amsmath, amssymb, nofootinbib]{revtex4}

\usepackage{graphicx}
\usepackage{color}
\usepackage{hyperref}
\usepackage{amsmath}
\usepackage{amssymb}
\usepackage[babel]{csquotes}
\usepackage{slashed}
\usepackage{braket, dsfont}
\usepackage{bm}

\begin{document}

\title{Hadronic vacuum polarization correction to atomic energy levels}

\author{S. Breidenbach}
\affiliation{Max Planck Institute for Nuclear Physics, Saupfercheckweg 1, 69117 Heidelberg, Germany}
\author{E. Dizer}
\affiliation{Max Planck Institute for Nuclear Physics, Saupfercheckweg 1, 69117 Heidelberg, Germany}
\author{H. Cakir}
\affiliation{Max Planck Institute for Nuclear Physics, Saupfercheckweg 1, 69117 Heidelberg, Germany}
\author{Z. Harman}
\email[]{harman@mpi-hd.mpg.de}
\affiliation{Max Planck Institute for Nuclear Physics, Saupfercheckweg 1, 69117 Heidelberg, Germany}

\date{\today}

\begin{abstract}

The shift of atomic energy levels due to hadronic vacuum polarization is evaluated in a semi-empirical way for hydrogenlike ions
and for muonic hydrogen. A parametric hadronic polarization function obtained from experimental cross sections of $e^- e^+$ annihilation into hadrons
is applied to derive an effective relativistic Uehling potential. The energy corrections originating from hadronic vacuum polarization are calculated
for low-lying levels using analytical Dirac-Coulomb wave functions, as well as bound wave functions accounting for the finite nuclear size.
Closed formulas for the hadronic Uehling potential of an extended nucleus as well as for the relativistic energy shift in case of a point-like nucleus
are derived. These results are compared to existing analytic formulas from non-relativistic theory.

\end{abstract}

\keywords{}

\maketitle

\section{Introduction}

The precision spectroscopy of hydrogen~\cite{Beyer2017,Udem2018,Fleurbaey2018,Bezginov2019}, hydrogenlike
and few-electron highly charged ions~\cite{Gillaspy2010,Marmar1986,Machado2018,Kubicek2014,Epp2007,Micke2020,Egl2019,Schmoeger2015,PhysRevLett.94.223001}
allows testing quantum electrodynamics (QED), a cornerstone of the Standard Model of particles and interactions, in unprecedented detail.
For example, two-loop effects and shifts due to nuclear structure have become accessible. At such precision, level shifts due to other forces need to
be considered as well. This holds especially true for muonic atoms, which recently became accessible by precision laser spectroscopy~\cite{Protonradius,Antognini2013}.
Therefore, in this article, the correction to the $1s$, $2s$ and $2p$ states of hydrogenlike ions due to virtual hadronic pair creation is studied.

\begin{figure}[h]
        \centering
        \includegraphics[scale=0.7]{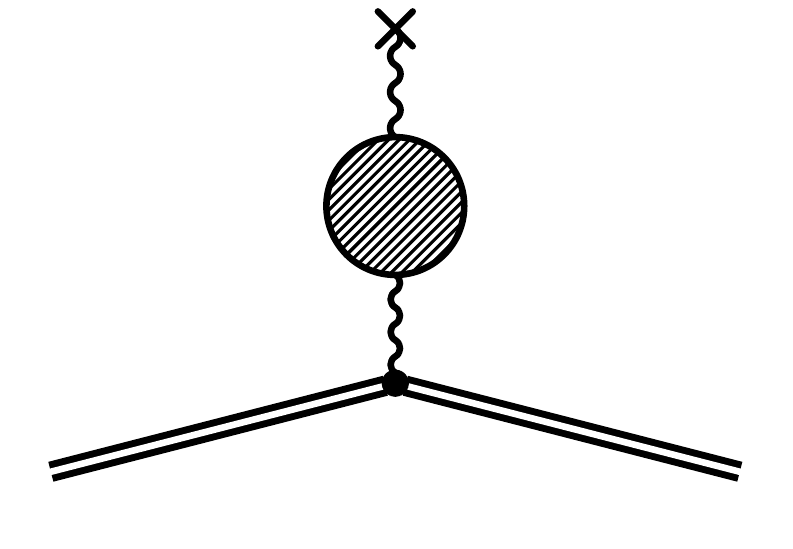}\\
        \caption{Feynman diagram depicting the leading hadronic vacuum polarization contribution. The double line represents
        a bound electron that interacts with the Coulomb field of a nucleus (the wavy line terminated by a cross) via a virtual photon,
        the propagator of which is modified by hadronic vacuum polarization represented by the shaded loop.}
\end{figure}

We investigate vacuum polarization (VP) corrections, whose largest contribution arises from virtual $e^- e^+$ pair creation.
This contribution is well understood and will only be mentioned here due to its importance and as reference for further corrections.
The next most important VP effect is due to virtual $\mu^- \mu^+$ pair creation, the contribution of which is
suppressed by the square of the electron-to-muon mass ratio~\cite{LandauLifshits4}, i.e. by a factor of $1/207^2 \approx 2 \times 10^{-5} $.
Apart from the different mass of the virtual fermions, the description of the muonic loop is equivalent to that of the electronic case.
The next one-loop contribution stems from several different virtual hadronic states, which call for a completely different description since the virtual
particles also interact via the strong interaction. First treatments were restricted to single hadrons, such as the $\rho$-meson \cite{Karshenboim1995},
one of the most important contributors to hadronic VP. Another approach is described in Ref.~\cite{Burkhardt1989}, in which the hadronic VP is
characterized by the total cross section of hadron production via $e^- e^+$ annihilation. 
Such experimental studies were largely motivated by the long-standing disagreement~\cite{muon-g-2021} of experiment and theory for the muon $g$ factor.
These discrepancies also triggered a range of perturbative and non-perturbative quantum chromodynamic calculations (see e.g.~\cite{Borsanyi2018,Blum2020} and references therein)
of hadronic vacuum polarization corrections. We employ a known parametric hadronic polarization
function for the photon propagator from \cite{Burkhardt2001} to account for the complete hadronic contribution in a semi-empirical manner. As we will see,
the high-energy part of the polarization function does not play a role when calculating shifts of atomic energy levels, therefore, perturbative quantum chromodynamic
results are not of relevance in our context.

An effective potential can be constructed from the parametrized VP function, called the hadronic Uehling potential.
The hadronic Uehling potentials of a point-like and a finite-sized nucleus are given analytically, and relativistic treatments are presented for both cases.
Subsequently, energy level shifts are calculated as a first-order perturbation employing the analytical Dirac-Coulomb wave function,
as well as with the numerically calculated wave function accounting for an extended nucleus. We note that such an approach assumes an infinitely heavy
nucleus, i.e. nuclear recoil effects are excluded in our treatment. These results are compared to the known non-relativistic
approximation for a point-like nucleus~\cite{Friar1999,Karshenboim1995,Karshenboim2021}. Results are given for a range of hydrogenlike systems from H to Cm$^{95+}$, and for muonic hydrogen.
The results for the different approaches will then be discussed in their uncertainty and applicability.

We use natural units with $\hbar=c=1$ for the reduced Planck constant and the speed of light, respectively,
and $\alpha = {e^2}$, where $\alpha$ is the fine-structure constant and $e$ is the elementary charge.
Three-vectors are denoted by bold letters. For brevity, we use the potential energy function $\delta V$ due to the Uehling potential
and refer to this as the Uehling potential, even though strictly speaking, it is the potential multiplied by the elementary charge.

\section{Vacuum polarization effects}

The interaction of a photon with virtual charged particles leads to a modification of its propagator.
This modified propagator can be described by the vacuum polarization tensor $\Pi_{\lambda \sigma}(q)$ and can be written as (see e.g.~\cite{Greiner7})
\begin{align}
i D^{\text{mod}}_{\mu \nu}(q) = i D_{\mu \nu}(q) + i D_{\mu \lambda}(q) \frac{i \Pi^{\lambda \sigma}(q)}{4 \pi} D_{\sigma \nu}(q) \, ,
\end{align}
with the unperturbed photon propagator $D_{\mu \nu}(q)$ and the four-momentum transfer $q$. Due to Lorentz and gauge invariance,
the polarization tensor can be cast into the form \cite{Greiner7}
\begin{align}
\Pi_{\lambda \sigma}(q) = (q^2 \eta_{\lambda \sigma} - q_\lambda q_\sigma) \Pi(q^2) \, ,
\end{align}
where $\eta_{\lambda \sigma}$ is the metric tensor [with diagonal elements $\left( 1,-1,-1,-1\right)$] and $\Pi (q^2)$ is the polarization function, which is divergent.
After regularization and charge renormalization, the divergent part of $\Pi (q^2)$ is isolated and only the regular part $\Pi^{\text{R}}(q^2)$
enters into physical calculations.

These leading vacuum polarization effects modify a static nuclear potential by the Uehling potential \cite{Greiner7}
\begin{align}
\delta V(\bm{x}) = \int \frac{d^3 q}{(2 \pi)^3} e^{i \bm{q} \cdot \bm{x}} \left( - \frac{4 \pi e}{\bm{q}^2} \right) \tilde{\rho}(\bm{q}) \, \text{Re} \left[\Pi^\text{R} \left(- \bm{q \,}^2 \right) \right] \, ,
\label{Uehling}
\end{align}
with the Fourier transform of the nuclear charge distribution $\tilde{\rho}(\bm{q})$ which is normalized to $Z e$.
Nuclear recoil corrections to vacuum polarization are not accounted for in such an effective potential approach by construction.
For a spherically symmetric nuclear charge distribution $\rho (r)$, the angular integration in Eq.~\eqref{Uehling} can be carried out, yielding
\begin{align}
\delta V(r) = - \frac{2 e}{\pi} \int_0^\infty d q \, j_0(q r) \tilde{\rho}(q) \, \text{Re} \left[\Pi^\text{R} \left(- q^2 \right) \right] \, ,
\label{Uehlingradial}
\end{align}
with the spherical Bessel function of 1st kind $j_k(x)$ of order $k$, and setting $|\bm{q}|=q$ and $|\bm{r}|=r$ from now on. The Uehling potential leads
to the leading perturbative shift of atomic energy levels, given by
\begin{align}
\Delta E_{n \kappa m} = & \bra{{n \kappa m}} \, \delta V \ket{{n \kappa m}} \notag \\
= & \int_{0}^{\infty} d r \, \delta V(r) \big( g^2_{n \kappa}(r) + f^2_{n \kappa}(r) \big) r^2 \,,
\label{pt1}
\end{align}
where $n$ is the principal quantum number, $\kappa$ is the relativistic angular momentum quantum number and $m$ is the magnetic quantum number.
The functions $g_{n \kappa}(r)$ and $f_{n \kappa}(r)$ are the large and small radial components, respectively, of the relativistic bound wave function
in the coordinate representation:
\begin{equation}
  \psi_{n \kappa m}(\bm{r}) = \langle {\bm{r}} | {{n \kappa m}}	 \rangle
  = 
  \begin{pmatrix} 
    g_{n\kappa}(r) \Omega_{\kappa m}(\bm{r}/r) \\
    i f_{n\kappa}(r) \Omega_{-\kappa m}(\bm{r}/r)
  \end{pmatrix} \,,
\end{equation}
where the $\Omega_{\kappa m}(\bm{r}/r)$ are spherical spinors~\cite{LandauLifshits4}.

\subsection{Hadronic vacuum polarization}

The leptonic polarization function is known analytically, and the corresponding VP shift can be calculated analytically, as an expansion in powers of the nuclear
coupling strength $Z \alpha$ or in certain cases even exactly. However, in the case of hadronic VP, the produced particles are strongly interacting, and a
perturbative quantum chromodynamic approach fails~\cite{Jegerlehner2002}.
One possibility is a semi-empirical approach to construct $\text{Re} \left[\Pi_{\text{had}}^\text{R} \left(q^2 \right) \right]$ via experimental
$e^-e^+ \rightarrow \text{hadrons}$ collision data \cite{Burkhardt1989}. The approach is summarized e.g. in~\cite{Jegerlehner2002,Breidenbach2018}.
The main steps are: The Kramers-Kronig relation enables to express the real part of a complex polarization function in terms of its imaginary part.
Then the optical theorem links a measurable total cross section $\sigma_{e^-e^+ \rightarrow \text{hadrons}}$ to the forward scattering amplitude,
in this case the imaginary part of the VP function. As a result, the cross section of the hadrons created in the pair annihilation process
enables the construction of the hadronic polarization function. This was performed e.g. in~\cite{Burkhardt1989}, where data from different experiments
and center-of-mass collision energy regions were compiled to yield an approximate parameterization of the polarization function:
\begin{align}
\text{Re} \left[\Pi_\text{had}^\text{R} \left(q^2 \right) \right] = A_i + B_i \ln \left(1 + C_i \left|q^2 \right| \right) \, ,
\label{parameterization}
\end{align}
with the constants $A_i$, $B_i$, $C_i$, which are given for different regions of $q^2$. For our evaluation, an updated version of this parameterization
with more energy regions will be used, as given in \cite{Burkhardt2001}. The parameters are shown in Table~\ref{tab:param} for completeness.

\begin{table}[t]
	\centering
	\begin{ruledtabular}
	\begin{tabular}{r l c c c}
		Region & Range [GeV] & $A_i$ & $B_i$ & $C_i$ [GeV$^{-2}$]\\
		\hline
		$k_0-k_1$ & $0.0-0.7$ & $0.0$ & $0.0023092$ & $3.9925370$\\
		$k_1-k_2$ & $0.7-2.0$ & $0.0$ & $0.0022333$ & $4.2191779$\\
		$k_2-k_3$ & $2.0-4.0$ & $0.0$ & $0.0024402$ & $3.2496684$\\
		$k_3-k_4$ & $4.0 - 10.0$ & $0.0$ & $0.0027340$ & $2.0995092$\\
		$k_4-k_5$ & $10.0-m_Z$ & $0.0010485$ & $0.0029431$ & $1.0$\\
		$k_5-k_6$ & $m_Z-10^4$ & $0.0012234$ & $0.0029237$ & $1.0$\\
		$k_6-k_7$ & $10^4-10^5$ & $0.0016894$ & $0.0028984$ & $1.0$\\
	\end{tabular}
	\end{ruledtabular}
	\caption{\label{tab:param}
        Values for the parameterization of the hadronic polarization function in Eq.~\eqref{parameterization} as given in Ref.~\cite{Burkhardt2001},
        with the mass of the $Z$ boson $m_Z$.}
\end{table}

\begin{figure}
	\centering
	\includegraphics[width=0.99\linewidth]{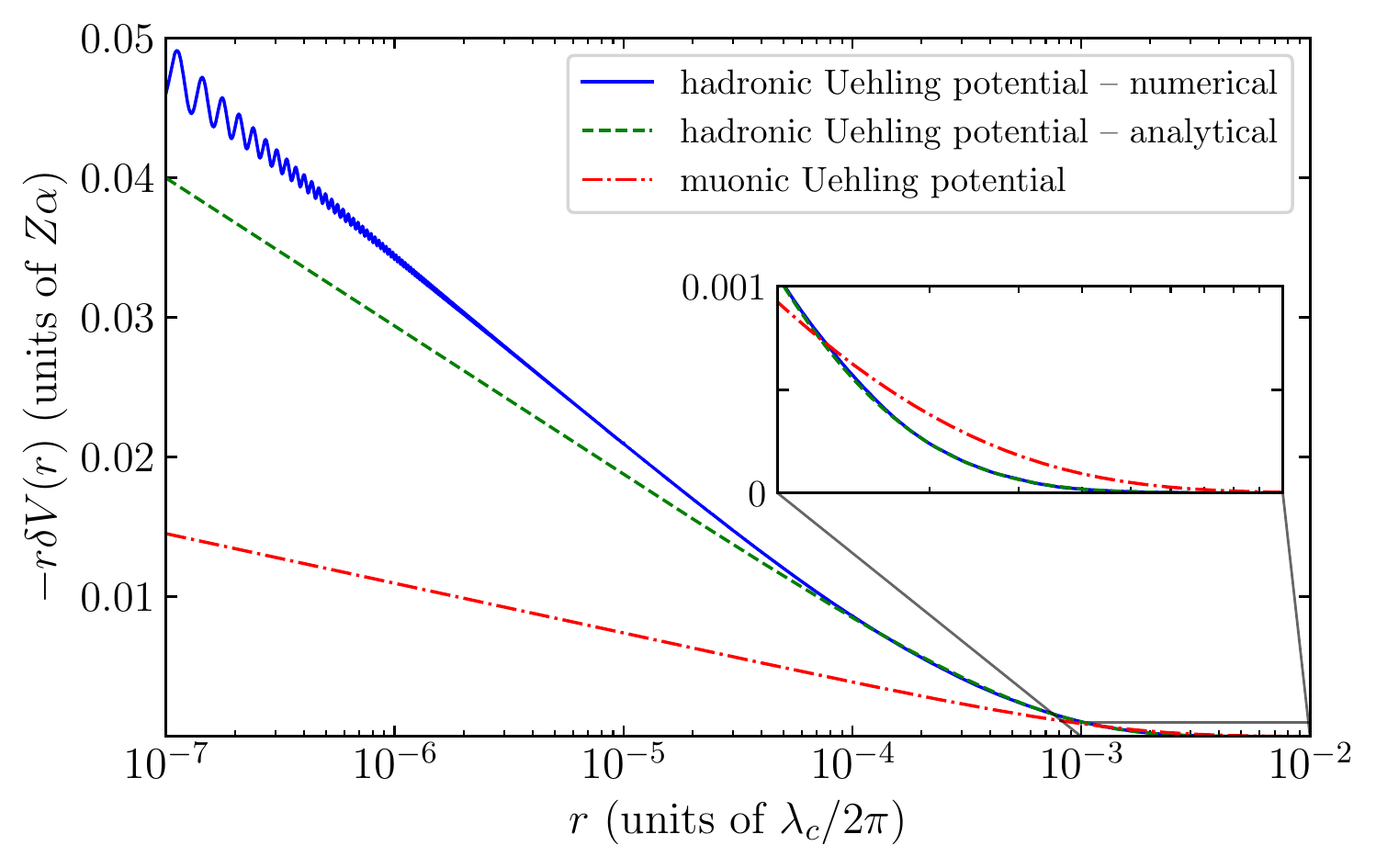}
	\caption{The numerical \eqref{fullUehlingpotential} and analytical \eqref{approximatedUehling} hadronic Uehling potential compared to the muonic Uehling potential as a function of the radius,
	in units of the reduced Compton wavelength $\lambda_{c}/2\pi$ of the electron (adapted from Ref.~\cite{Dizer2020}).}
	\label{fig:potential}
\end{figure}

The Uehling potential for this parameterization, assuming a spherically symmetric proton distribution, is therefore given by
\begin{align} \label{fullUehlingpotential}
	\delta V^{\text{full}}_{\text{fns}}(r) = - \frac{2 e}{\pi} \sum_{i=1}^{7} \bigg[ & \int_{k_{i-1}}^{k_{i}}d q \, j_0(q r) \tilde{\rho}(q) \\
 	& \times \big[ A_i + B_i \ln(1 + C_i q^2) \big] \bigg] \notag \, .
\end{align}
For our purposes, a good approximation for the full polarization function is to use the parameters of its first momentum region up to infinity, i.e. only using
the parameters $A_1$, $B_1$ and $C_1$. In this case, the Uehling potential of a point-like nucleus ($\tilde{\rho}(\bm{q}) = Z e$) simplifies to
\begin{align} \label{approximatedUehling}
	\delta V^{\text{approx}}_{\text{point}}(r) & = -  \frac{2 Z \alpha}{\pi} \int_{0}^{\infty}d q \, j_0(q r) \big[B_1 \ln \left(1 + C_1 q^2 \right) \big] \notag \\
	& = - \frac{2 Z \alpha}{r} B_1 E_1\left( \frac{r}{\sqrt{C_1}} \right)\,,
\end{align}
with the exponential integral
\begin{align}
	E_n(x) = \int_1^{\infty} dt \, \frac{e^{-xt}}{t^n} \,.
\end{align}
This approximation is physically well motivated because the low-energy region is the most important one in atomic physics, and the original range of applicability
for the parameters at $0.7$ GeV should be sufficient for our applications. In fact we will show in Section~\ref{sec:Results} that at least up to $Z=96$,
no difference between this approximation and the full numerical result is observable for the calculated energy shifts within our level of uncertainty.
The analytical approximation reduces numerical errors and speeds up the calculations. The analytical and numerical hadronic Uehling potentials of a point-like
nucleus are displayed in Fig.~\ref{fig:potential} and are compared to the well-known muonic Uehling potential. The oscillations at low distances
in the potential defined by Eq.~(\ref{fullUehlingpotential}) are due to the upper momentum cut-off in the parameterized polarization function, and therefore not
physical.

The energy shift for the $1s$ state and a point-like nucleus in first-order perturbation theory is given by the expectation value~\cite{Dizer2020}
\begin{align}
	&\Delta E_{\text{rel., point}}^{\text{analytical}}(1s) =  \langle {{1s}} \vert \delta V^{\text{approx}}_{\text{point}} \vert {{1s}} \rangle \notag \\
	&= - \frac{Z\alpha \lambda (2 \lambda \sqrt{C_1} )^{2 \gamma} B_1}{\gamma^2} {}_2F_1\left(2\gamma, 2\gamma; 1+2\gamma;-2 \lambda \sqrt{C_1}\right) \,,
\end{align}
using the analytical Coulomb-Dirac wave function~\cite{LandauLifshits4} with the charge number $Z$, the hypergeometric function $_2F_1\left(a, b; c; z\right)$ and
\begin{align}
\lambda = Z \alpha m_{\rm e}  , \ \ \ \gamma = \sqrt{1 - (Z \alpha)^2} \, .
\end{align}
The Taylor expansion of this all-order result up to order $(Z\alpha)^6$ is
\begin{align}\label{energyshiftpointexpansion1s}
\Delta E_{\text{rel., point}}^{\text{analytical}}(1s) = &- 4 B_1 C_1 m_{\rm e}^3(Z \alpha)^4  \notag \\
& + \frac{32 B_1 C_1^{3/2} m_{\rm e}^4 (Z \alpha)^5}{3} \notag \\
& - 4 B_1 C_1 m_{\rm e}^3 (Z \alpha)^6 \notag \\
& \times \left[ 1+6 C_1 m_{\rm e}^2 - \ln(2 Z\alpha\sqrt{C_1} m_{\rm e}) \right]  \notag \\
& + \dots\,.
\end{align}
A similar calculation may be performed for the $2s$ state, yielding approximately
\begin{align}\label{energyshiftpointexpansion2s}
\Delta E_{\text{rel., point}}^{\text{analytical}}(2s) = &- \frac{1}{2} B_1 C_1 m_{\rm e}^3(Z \alpha)^4 \notag \\
&+ \frac{4 B_1 C_1^{3/2} m_{\rm e}^4 (Z \alpha)^5}{3} + \dots .
\end{align}
The very first terms of Eq.~(\ref{energyshiftpointexpansion1s}) and (\ref{energyshiftpointexpansion2s}) agree with the non-relativistic formula of Friar et al. \cite{Friar1999}
(see also Ref.~\cite{Karshenboim2021})
for the case of $n=1,2$, respectively.

While atomic wave functions with higher orbital angular momenta have a small overlap with the short-distance region where the hadronic Uehling potential is significant,
for completeness, we also discuss the case of the $2p_{1/2}$ (i.e. total angular momentum $j=1/2$, $\kappa=1$) orbital. Expanding the resulting fully relativistic
formula in powers of  $Z\alpha$, we obtain
\begin{align}
\label{energyshiftpointexpansion2p1/2}
\Delta E_{\text{rel., point}}^{\text{analytical}}(2p_{1/2}) = &- \frac{B_1 C_1 (3 + 4 C_1 m_{\rm e}^2) m_{\rm e}^3}{32} (Z\alpha)^6 \notag \\
& + \frac{B_1 C_1^{3/2} (5 + 24 C_1 m_{\rm e}^2) m_{\rm e}^4}{60} (Z\alpha)^7 \notag \\
& + \dots\,.
\end{align}
Let us note that the leading term in $Z\alpha$ may be also obtained in a different way: To the lowest order,
the large and small radial components of the  Dirac wave function can be approximated as
\begin{align}
g_{2p_{1/2}}(r) &= - \left(\frac{Z\alpha}{2}\right)^{1/2} \frac{(Z\alpha)^2 r }{2\sqrt{3}} \exp\left(-\frac{Z\alpha r}{2}\right) , \\
f_{2p_{1/2}}(r) &= - \left(\frac{Z\alpha}{2}\right)^{3/2} \frac{3Z\alpha }{2\sqrt{3}} \exp\left(-\frac{Z\alpha r}{2}\right) .
\end{align}
(see e.g. \cite{Cakir2020}). By evaluating the integral~(\ref{pt1}) with these functions and with the approximate potential (\ref{approximatedUehling}) to the leading
order in $Z\alpha$, one can reproduce the term of order $(Z\alpha)^6$ in Eq.~(\ref{energyshiftpointexpansion2p1/2}).
Thus, the hadronic shift for this state is suppressed by a small factor of $(Z\alpha)^2$ compared to that of the $s$ state, and its value is negligible, as we will discuss
in Section~\ref{sec:Results}.

For completeness, for the $2p_{3/2}$ (i.e. total angular momentum $j = 3/2$, $\kappa = -2$) orbital, we get the following result:
\begin{align}
\label{energyshiftpointexpansion2p3/2}
\Delta E_{\text{rel., point}}^{\text{analytical}}(2p_{3/2}) = &- \frac{B_1 C_1^2 m_{\text{e}}^5}{8} (Z\alpha)^6 \notag \\
& + \frac{2 B_1 C_1^{5/2} m_{\text{e}}^6}{5} (Z\alpha)^7 \notag \\
& + \dots\,.
\end{align}
Interestingly, the expansion coefficients for this orbital also appear in the expansion for the $2p_{1/2}$ state, see Eq.~\eqref{energyshiftpointexpansion2p1/2}.

\subsection{Finite-Size Hadronic Uehling Potential}

One possibility to model a finite-size nucleus is employing a spherical homogeneous charge distribution $\rho(r)$ with the effective radius~$R$,
\begin{align}
	&\rho(r) = \frac{3 Z e}{4 \pi R^3} \theta(R-r)\,,
	\label{sphericaldistribution}
\end{align}
for which the momentum representation $\tilde{\rho} \left(q \right)$ can be easily found:
\begin{align}
	&\tilde{\rho} \left(q \right) = Z e \frac{3 j_1(q R)}{q R} \,,
\end{align}
where the radius $R$ is related to the root-mean-square (rms) nuclear radius via $R = \sqrt{5/3 \langle r^2 \rangle}$.
To calculate the Uehling potential corresponding to this charge distribution one substitutes $\tilde{\rho} \left(q \right)$ into Eq.~\eqref{fullUehlingpotential}.

Alternatively to Eq.~\eqref{fullUehlingpotential}, the Uehling potential corresponding to a finite-size nucleus can also be calculated by convoluting the Uehling potential
of a point-like nucleus with a charge distribution $\rho(\bm{x})$ in real space:
\begin{align}\label{convolution}
\delta V^{\text{approx}}_{\text{fns}}(\bm{r}) = & \frac{1}{Ze}\int d^3x \ \rho(\bm{x}) \, \delta V^{\text{approx}}_{\text{point}}(\bm{r}-\bm{x}) \, .
\end{align}
Using our approximated Uehling potential for a point-like nucleus from Eq.~\eqref{approximatedUehling} and a spherically symmetric charge distribution, the formula simplifies to
\begin{align}\label{convolutionradialshort}
\delta V^{\text{approx}}_{\text{fns}}(r) = & - \frac{4 \pi e B_1 \sqrt{C_1}}{r} \int_0^{\infty} dx \ x \rho(x) D^{-}_2(r,x) \, ,
\end{align}
with
\begin{align}
	D^{\pm}_n(r,x) = E_n \left(\frac{|r-x|}{\sqrt{C_1}} \right) \pm E_n\left(\frac{|r+x|}{\sqrt{C_1}} \right) \,.
\end{align}
This integral can be solved analytically for the homogeneously charged model, divided into two separate solutions for the regions outside and inside of the nucleus: \\
$r > R$:
\begin{align}
\delta V^{\text{approx}}_{\text{fns,out}}(r)  = - & \frac{3 Z \alpha B_1 \sqrt{C_1}}{r R^3} \notag \\
& \times \left\{  \sqrt{C_1} R \, D^{+}_3(r,R) - C_1 D^{-}_4(r,R) \right\}.
\end{align}
$r \leq R$:
\begin{align}
\delta V & ^{\text{approx}}_{\text{fns,in}}(r) = - \frac{3 Z \alpha B_1 \sqrt{C_1}}{r R^3} \notag \\ 
&\times \left\{  \sqrt{C_1} r + \sqrt{C_1} R E_3\left( \frac{r+R}{\sqrt{C_1}} \right) + C_1 E_4\left( \frac{r+R}{\sqrt{C_1}} \right) \right. \notag \\
&- \frac{1}{6} e^{\frac{r-R}{\sqrt{C_1}}} \left(2C_1 + \sqrt{C_1}(r+2R)+(r-R)(r+2R)\right) \notag \\
&- \left. \frac{(r-R)^2(r+2R)}{6\sqrt{C_1}}E_1\left( \frac{R-r}{\sqrt{C_1}} \right)  \right\}.
\end{align}

\section{Results}
\label{sec:Results}

In order to have a reference for the other approaches, we calculate the hadronic energy shift in the non-relativistic approximation
$\Delta E_{\text{non-rel., point}}^{\text{analytical}}$, which corresponds to the first term from Eq.~\eqref{energyshiftpointexpansion1s}.
The result for hydrogen is
\begin{align}
\Delta E_{\text{non-rel., point}}^{\text{had. VP}}(1s) & = -1.395(17) \times 10^{-11} \text{ eV} \notag \\
& = 0.6647(81)\, \Delta E_{\text{non-rel., point}}^{\text{muonic VP}}(1s) \, ,
\end{align}
with the energy shift due to muonic VP denoted by $\Delta E_{\text{non-rel.}}^{\text{muonic VP}}$. This is in good agreement with the formula
\begin{align}
\Delta E_{\text{non-rel.}}^{\text{had. VP}} = 0.671(15) \Delta E_{\text{non-rel.}}^{\text{muonic VP}}
\end{align}
from \cite{Friar1999}, with the difference stemming from using more recent experimental constants $B_1$ and $C_1$ in Eq. (\ref{energyshiftpointexpansion1s})
as compared to \cite{Friar1999}.
For regular and muonic H, our results (see Tables~\ref{tab:results}, \ref{tab:excresults}, \ref{tab:muonichydrogen}) within this model agree well with the
recent results of Ref.~\cite{Karshenboim2021} (see Table~2 therein).

\begin{table*}[ht!]
	\begin{ruledtabular}
		\begin{tabular}{llllll}
			$Z$ & $R_{\text{rms}}$ [fm] & $\Delta E_{\text{non-rel., point}}^{\text{analytical}}$ [eV] & $\Delta E_{\text{rel., point}}^{\text{analytical}}$ [eV] & $\Delta E_{\text{rel., fns}}^{\text{approx}}$ [eV] & $\Delta E_{\rm LS}$ [eV]\\ \hline
			$1$ & $0.8783(86)$  & $-1.395(17)[-11]$ & $-1.396(17)[-11]$ & $-1.396(17)[-11]$ & $3.3800262(7)(57)[-5]$ \\
			$14$ & $3.1224(24)$ & $-5.361(67)[-7]$  & $-5.918(73)[-7]$  & $-5.756(72)[-7]$  & $4.80447(18)(4)[-1]$   \\
			$20$ & $3.4776(19)$ & $-2.233(28)[-6]$  & $-2.713(33)[-6]$  & $-2.560(32)[-6]$  & $1.63263(6)(2)[0]$     \\
			$36$ & $4.1884(22)$ & $-2.344(29)[-5]$  & $-4.270(50)[-5]$  & $-3.485(43)[-5]$  & $1.18259(16)(3)[1]$    \\
			$54$ & $4.7859(48)$ & $-1.187(15)[-4]$  & $-4.445(48)[-4]$  & $-2.706(34)[-4]$  & $4.6920(18)(6)[1]$     \\
			$74$ & $5.3658(23)$ & $-4.184(52)[-4]$  & $-5.098(46)[-3]$  & $-1.801(22)[-3]$  & $1.5422(13)(2)[2]$     \\
			$82$ & $5.5012(13)$ & $-6.309(79)[-4]$  & $-1.413(11)[-2]$  & $-3.693(46)[-3]$  & $2.4440(26)(3)[2]$     \\
		\end{tabular}
		\caption{\label{tab:results} Hadronic vacuum polarization energy shifts using different approaches for the $1s$ ground state of the considered hydrogenic
			systems: the non-relativistic approximation $\Delta E_{\text{non-rel., point}}^{\text{analytical}}$, the relativistic analytical formula for a
			point-like nucleus $\Delta E_{\text{rel., point}}^{\text{analytical}}$, and the analytical finite-size Uehling potential with numerical
			finite-size wave functions $\Delta E_{\text{rel., fns}}^{\text{approx}}$. Powers of 10 are enclosed in brackets, and uncertainties are indicated in parentheses.
			The root-mean-square nuclear charge radii in the second column are taken from \cite{Angeli2013}.
			The last column shows for comparison the total Lamb shift contribution $\Delta E_
			{\rm LS}$ from Ref.~\cite{Yerokhin2015}.
			The values have two uncertainties given in round brackets:
			the second one is due to the error of the nuclear charge radius, whereas the first one represents 
			all other errors of individual theoretical contributions added quadratically.
		}
		\label{results}
	\end{ruledtabular}
\end{table*}

\begin{table*}[tbh]
	\begin{ruledtabular}
		\begin{tabular}{lllllll}
			$Z$  &  $\Delta E_{\text{rel., fns}}^{\text{approx}}(2s)$ [eV] & $\Delta E_{\rm LS}(2s)$ [eV] & $\Delta E_{\text{rel., fns}}^{\text{approx}}(2p_{1/2})$ [eV] & $\Delta E_{\rm LS}(2p_{1/2})$ [eV] & $\Delta E_{\text{rel., fns}}^{\text{approx}}(2p_{3/2})$ [eV] & $\Delta E_{\rm LS}(2p_{3/2})$ [eV]  \\
			\hline
			$1$  &  $-1.745(22)[-12]$ & $4.3218005(8)(72)[-6]$ &  $-1.743(22)[-17]$ & $-5.30919(4)(0)[-8]$  & $-6.427(80)[-23]$ & $5.177459[-8]$      \\
			$14$ &  $-7.262(91)[-8]$  & $6.40329(23)(5)[-2]$   &  $-1.431(19)[-10]$ & $-1.7316(4)(0) [-3]$  & $-4.168(52)[-15]$ & $2.1808(4)(0)[-3]$  \\
			$20$ &  $-3.260(41)[-7]$  & $2.21409(9)(2)[-1]$    &  $-1.321(17)[-9]$  & $-6.2940(35)(0)[-3]$  & $-4.535(57)[-14]$ & $9.6566(34)(0)[-3]$ \\
			$36$ &  $-4.631(58)[-6]$  & $1.68814(25)(4)[0]$    &  $-6.261(78)[-8]$  & $-3.4426(62)(1)[-2]$  & $-2.602(33)[-12]$ & $1.2089(9)(0)[-1]$  \\
			$54$ &  $-3.887(49)[-5]$  & $7.1723(27)(9)[0]$     &  $-1.251(16)[-6]$  & $6.0317(72)(36)[-1]$  & $-5.036(63)[-11]$ & $7.413(10)(0)[-1]$  \\
			$74$ &  $-2.932(37)[-4]$  & $2.5876(20)(4)[1]$     &  $-1.957(24)[-5]$  & $1.6390(33)(3)[0]$    & $-6.217(78)[-10]$ & $3.1615(30)(0)[0]$  \\
			$82$ &  $-6.403(80)[-4]$  & $4.2924(44)(4)[1]$     &  $-5.541(69)[-5]$  & $3.9045(72)(4)[0]$    & $-1.462(18)[-9]$  & $5.1088(57)(0)[0]$  \\
		\end{tabular}
		\caption{\label{tab:excresults} Hadronic vacuum polarization energy shifts for the $2s$ and $2p$ excited states of the considered hydrogenic
		systems. Notations and nuclear radii used are as in Table~\ref{tab:results}.
		}
	\label{results}
	\end{ruledtabular}
\end{table*}

The values for the hadronic energy shift with an extended nucleus were calculated numerically using two different methods,
both yielding the same results. The first method consists of solving the Dirac equation, with inclusion of the potentials
for an extended nucleus, using a B-splines representation and extracting the corresponding energy eigenvalues. As a consistency
check, these results were reproduced by calculating the expectation value of the FNS hadronic Uehling potential with respect to
the semi-analytic wave functions belonging to a spherical nucleus given in \cite{Patoary2018}.  The results for the ground state of hydrogenlike systems
H, Si, Ca, Xe, Kr, W, Pb and Cm are shown, and different methods of approximation are contrasted in Table \ref{tab:results}.
Results for $n=2$ excited states are presented in Table \ref{tab:excresults}, while values for muonic hydrogen are given in Table \ref{tab:muonichydrogen}.

The errors given in Tables~\ref{tab:results} and \ref{tab:excresults} are based on the numerical convergence of the results, the uncertainty
of the nuclear root-mean-square radii and, dominantly, the difference with respect to values obtained by using another set of parameters to describe the
polarization function, stemming from \cite{Burkhardt1995}. Numeric values using the approximated and the full Uehling potential always match very well
within the uncertainty given, with the exception of hydrogen due to numerical difficulties in the evaluation of the full Uehling potential. In particular
for this case, the result from the approximated analytical formula should be correct due to the low $Z$ value. \\

In order to show the range of
validity for our approximation, we computed the energy shift for $Z=96$ with $R_{\text{rms}}=5.85$ fm. The approximated and full Uehling potential
evaluated with both numerical methods yield all the same mean value: $-1.2637 \times 10^{-2}$ eV. We conclude that the approximated
analytical Uehling potential incorporates all relevant information at least up to $Z=96$ and is therefore applicable in all practical computations.
Its use also reduces the numerical errors and speeds up the calculations significantly, thus rendering it also the method of choice for further applications. \\

\begin{table}[t]
	\centering
	\begin{ruledtabular}
		\begin{tabular}{lll}
			State      & $\Delta E_{\text{non-rel., point}}^{\text{analytical}}$ [meV] & $\Delta E_{\text{rel., fns}}^{\text{full}}$ [meV] \\
			\hline
			$1s$       & $-1.234(15)[-1]$    & $-1.229(15)[-1]$ \\
			$2s$       & $-1.542(19)[-2]$    & $-1.53(5)[-2]$   \\
			$2p_{1/2}$ & $- 1.631(22)[-7]$   & $-1.8(1)[-7]$    \\
		\end{tabular}
	\end{ruledtabular}
	\caption{Results for muonic hydrogen within the non-relativistic approach
	$\Delta E_{\text{non-rel., point}}^{\text{analytical}}$, i.e. using the analytical formulas to the lowest order in $Z\alpha$,
	and employing the analytical finite-size Uehling potential with numerical finite-size wave functions $\Delta E_{\text{rel., fns}}^{\text{full}}$.
	In both columns, nuclear recoil effects are excluded.}
	\label{tab:muonichydrogen}
\end{table}

We also observe that, except for hydrogen, the point-like nucleus values all differ significantly from the finite-size values. We conclude that one
should always include the effects of a finite-size nucleus in a relativistic approach. In order to estimate the error stemming from the assumed
nuclear model, we solved Eq.~\eqref{convolutionradialshort} for $Z=82$ with the nuclear charge density modeled by a Fermi distribution with a
skin thickness of $2.3$~fm. The result for the perturbative energy shift, $-3.646 \times 10^{-3}$~eV, differs from the result assuming a homogeneous
nuclear charge distribution on the 1\% level and is therefore negligible.

The highest hadronic VP energy shift for the ions considered is on the meV level; this is the case for the very heavy element Pb.
Such a small effect can not be resolved yet experimentally in a $K\alpha$ x-ray transition (see e.g. \cite{PhysRevLett.94.223001,Kubicek2014}).
Furthermore, the theoretical Lamb shift values and their uncertainties~\cite{Yerokhin2015} given for comparison in
Tables~\ref{tab:results} and \ref{tab:excresults} show that uncertainties arising from the nuclear charge distribution need to be improved
by at least one order of magnitude, and QED terms need to be evaluated more accurately in future to render hadronic VP observable.
(For more recent Lamb shift results, see e.g. Refs.~\cite{Yerokhin2019,Yerokhin2018Two,Indelicato2019,Dorokhov2019,Shabaev2018,Czarnecki2016,Tupitsyn2016}.)

Therefore, we also consider another system that may feature measurable shifts,
namely, muonic hydrogen. In Table \ref{tab:muonichydrogen}, results for this system are shown, which can be simply obtained by replacing in
the above formulas $m_{\rm e}$ by the muon mass $m_{\mu}$ (note that generally, in our approach we neglect nuclear recoil effects, and the
use of the reduced mass $m_{\mu} m_{\rm p}/(m_{\mu}+ m_{\rm p})$, with $m_{\rm p}$ being the proton mass, would not be appropriate in a
relativistic theory).

We also list results for the $2s$ and $2p_{1/2}$ states, since these classical Lamb shift levels were involved in the
muonic hydrogen laser spectroscopic experiments determining the radius of the proton~\cite{Protonradius,Antognini2013}. The uncertainty of
the experimental muonic Lamb shift, $49881.88(76)$~GHz~\cite{Protonradius} (or, more recently, $49881.35(65)$~GHz~\cite{Antognini2013}) translates
to $0.003$~meV, which would be in principle sufficient to resolve the hadronic VP contribution, motivating an accurate evaluation of the latter.
However, currently the experimental value of the muonic hydrogen Lamb shift is limited by the uncertainty of the proton radius~\cite{Protonradius,Antognini2013}.

For the hadronic VP shift of the $2s$ energy level we obtain  a value of $-0.0153(5)$~meV including finite nuclear size effects. This result agrees
with the non-relativistic approach for a point-like nucleus due to the smallness of $Z \alpha$.  We note that in the non-relativistic theory, recoil effects
can be accounted for by replacing the muon mass $m_{\mu}$ with the reduced mass of the atom, reproducing the literature value~\cite{Eides2001,Borie2005}
of $-0.0112$~meV (or the most recent result of $-0.01116(7)$~meV of Ref.~\cite{Karshenboim2021}) for the hadronic shift of the $2s$ state. As the Table also shows, the hadronic VP correction to the $2p_{1/2}$ energy level is negligible
at the current level of experimental and theoretical uncertainties for muonic hydrogen.

\section{Summary}

The rising precision of experimental spectroscopic measurements and theoretical predictions calls for more detailed description of known effects.
The muonic VP is already an established part of theory \cite{Franosch1991}.
In order to assure that the hadronic VP does not limit the precision of theory, a generalized approach is desirable. In this paper we take into account
relativistic effects and finite nuclear size effects, which are relevant in highly charged ions. Therefore, this paper is a
contribution to understand and diminish the theoretical uncertainty induced by hadronic vacuum polarization in precision spectroscopy.

In this work, an effective potential was constructed by using a parameterized hadronic function
obtained from experimental data. An analytic formula for the finite-size Uehling potential was found, and it was shown to agree with the numerical approach
in all examined systems. Finally, energy shifts induced by the hadronic Uehling potential were computed, including an analytical relativistic formula and
two different numerical methods. We would like to note that hadronic VP diagrams, in which the nucleus interacts strongly with the loop hadrons,
are not accounted for by this approach, nor are hadronic virtual light-by-light scattering effects. The energy shifts were determined for hydrogenlike
systems ranging from $Z=1$ up to $Z=96$, and for muonic hydrogen. The results for the energy shift induced by hadronic VP exhibit that for our desired
level of accuracy, it is sufficient to describe hydrogen and light ions non-relativistically, and heavier systems relativistically, using the analytical finite-size
Uehling potential.
The main source of uncertainty is expected to stem from the applied nuclear model. This can be improved by using more elaborate
charge distribution models~\cite{Valuev2020} in Eq.~\eqref{convolution}, respectively Eq.~\eqref{Uehling}, and taking into account relativistic nuclear recoil effects.
Another main source of error is due to uncertainties in the parameterization of the empirical hadronic VP function. An advanced parameterization,
especially in the low-energy region, could improve this area of precision science.

Nowadays, besides energies of transitions between atomic levels, the $g$ factors of few-electron ions can be experimentally determined to
high precision by means of the continuous Stern-Gerlach effect in Penning trap setups~\cite{Sturm2013,Sturm2014,Verdu2004,Arapoglou2019,Quint2001}. This motivates
the extension of the calculation of hadronic vacuum polarization corrections to the bound-electron $g$ factor. Such calculations are currently underway.

\begin{acknowledgements}

S.B. would like to thank the people of the Max Planck Institute for Nuclear Physics, especially the theory division lead by Christoph H. Keitel for the hospitality during the work.
This publication was also supported by the Collaborative Research Centre 1225 funded by Deutsche Forschungsgemeinschaft (DFG, German Research Foundation)
-- Project-ID 273811115 -- SFB 1225.

\end{acknowledgements}

\end{document}